\documentstyle[preprint,aps]{revtex}

\input psfig.sty
\tightenlines 

\begin{document}
\draft
\title{Direct Urca processes on nucleons in cooling neutron stars}
\author{L. B. Leinson}
\address{Institute of Terrestrial Magnetism, Ionosphere and Radio Wave\\
Propagation RAS, 142190 Troitsk, Moscow Region, Russia\\
E-mail: leinson@izmiran.rssi.ru}
\maketitle

\begin{abstract}
We use the field theoretical model to perform relativistic calculations of
neutrino energy losses caused by the direct Urca processes on nucleons in
the degenerate baryon matter. By our analysis, the direct neutron decay in
the superdense nuclear matter under beta equilibrium is open only due to the
isovector meson fields, which create a large energy gap between protons and
neutrons in the medium. Our expression for the neutrino energy losses,
obtained in the mean field approximation, incorporates the effects of
nucleon recoil, parity violation, weak magnetism, and pseudoscalar
interaction. For numerical testing of our formula, we use a self-consistent
relativistic model of the multicomponent baryon matter. The relativistic
emissivity of the direct Urca reactions is found substantially larger than
predicted in the non-relativistic approach. We found that, due to weak
magnetism effects, relativistic emissivities increase by approximately
40-50\%, while the pseudoscalar interaction only slightly suppresses the
energy losses, approximately by 5\%.
\end{abstract}


\widetext

\section{Introduction}

Modern calculations \cite{PBPELK97} based on relativistic equations of state
indicate that the neutron star cores consist of neutrons with the admixture
of protons, electrons, muons and some exotic particles (including hyperons,
K-mesons, quarks and so on\ldots ). The composition is governed by the
charge neutrality and equilibrium of the medium under weak processes $B_{%
{\rm 1}}\rightarrow B_{{\rm 2}}+l+\bar{\nu}_{l}$, $\ \ \ B_{{\rm 2}%
}+l\rightarrow B_{{\rm 1}}+\nu _{l}$, where $B_{{\rm 1}}$ and $B_{{\rm 2}}$
are baryons (or quarks), and $l$ is a lepton, either an electron or a muon.
These reactions, widely known as the direct Urca processes, are a central
point of any modern scenarios of evolution of neutron stars. Neutrino energy
losses caused by the direct Urca processes lead to a rapid cooling of
degenerate neutron star cores \cite{LRP94}. The corresponding reactions on
nucleons, $n\rightarrow p+l+\bar{\nu}_{l}$, $\ \ \ p+l\rightarrow n+\nu _{l}$%
, are the most powerful sources of neutrinos and antineutrinos in cooling
neutron stars. In spite of widely adopted importance of these reactions, the
corresponding neutrino energy losses are not well investigated yet. A simple
formula suggested by Lattimer et al. \cite{Lat91} more than ten years ago
has been derived in a non-relativistic manner. Actually, however, the
superthreshold proton fraction, necessary for the direct Urca processes to
operate in the degenerate nuclear matter, appears at large densities, where
the Fermi momenta of participating nucleons are comparable with their
effective mass. Moreover, according to modern numerical simulations, the
central density of the star can be up to eight times larger than the nuclear
saturation density \cite{PBPELK97}. This implies a substantially
relativistic motion of nucleons in the superdense neutron star core. The
appropriate equation of state for such a matter is actually derived in the
relativistic approach, and the relevant neutrino energy losses must be
consistent with the used relativistic equation of state. Some aspects of
this problem was studied by Leinson and P\'{e}rez \cite{PLB}, who have
estimated relativistic effects of baryon recoil and parity violation in the
direct Urca processes including also effects of the baryon mass difference.
This approach is useful, for example, for the direct Urca reactions changing
the baryon strangeness because, in this case, the contribution of the weak
magnetism into the matrix element of the beta-decay is not well known even
for free particles. For nucleons, the relativistic regime should incorporate
the effects of weak magnetism and pseudoscalar interaction, which
drastically influence the neutrino emissivity of the corresponding
reactions. To demonstrate this, in the present paper we consider the totally
relativistic direct Urca process on nucleons. We utilize the Walecka-type
relativistic model of baryon matter \cite{Serot}, where the baryons interact
via exchange of $\sigma $, $\omega $, and $\rho $ mesons, and perform the
calculation of the neutrino energy losses in the mean field approximation.
This approximation is widely used in the theory of relativistic nuclear
matter, and allows to calculate in a self-consistent way the composition of
the matter together with energies, and effective masses of the baryons. In
Section II we begin with considering the relativistic kinematics of the
neutron beta decay in the medium under beta equilibrium. We consider a free
gas model and some field theoretical models of nuclear matter to demonstrate
that the direct neutron decay in neutron stars is open only due to strong
interactions caused by isovector mesons. We shortly discuss the energies and
the wave functions of nucleons in the mean field approximation and
demonstrate the nonconservation of the charged vector current of nucleons in
this model. The matrix element of the neutron beta decay is derived in
Section III. In Section IV we calculate the neutrino energy losses caused by
the direct Urca on nucleons in the degenerate nuclear matter under chemical
and thermal equilibrium. In Section V we inspect the non-relativistic limit
of the neutrino energy losses in order to compare this with the expressions
earlier obtained in \cite{Lat91} and \cite{PLB}. Efficiency of the
relativistic approach is numerically studied in Section VI. We evaluate the
neutrino energy losses due to the direct Urca processes on nucleons in the
multicomponent baryon matter under beta equilibrium and compare the
relativistic result with that predicted by the known non-relativistic
formula. We specially discuss the contributions of weak magnetism and
pseudoscalar interaction. Summary and conclusion are in Section VII. In
Appendix, we discuss some details of the model used for the nuclear matter.

In what follows we use the system of units $\hbar =c=1$ and the Boltzmann
constant $k_{B}=1$. Summation over repeated Greek indexes is assumed

\section{Kinematics of the reaction and models of nuclar matter.}

Before calculating the neutrino energy losses caused by the direct Urca
processes, let us examine relativistic kinematics of the reaction, $%
n\rightarrow p+l+\bar{\nu}_{l}$, in the degenerate matter under beta
equilibrium. In what follows we consider massless neutrinos of energy and
momentum $k_{1}=\left( \omega _{1},{\vec k}_{1}\right) $ with $\omega
_{1}=\left| {\vec k}_{1}\right| .$ The energy-momentum of the final lepton $%
l=e^{-},\mu ^{-}$ of mass $m_{l}$ is denoted as $k_{2}=\left( \omega _{2},%
{\vec k}_{2}\right) $ with $\omega _{2}=\sqrt{{\vec k}_{2}^{2}+m_{l}^{2}} $.
Thus, the energy and momentum conservation in the beta decay is given by the
following equations%
\begin{eqnarray}
E_{{\rm n}}\left( {\vec p}\right) -E_{{\rm p}}\left( {\vec p}^{\prime }\right)
-\omega _{1}-\omega _{2} &=&0  \nonumber \\
{\vec p}-{\vec p}^{\prime }-{\vec k}_{1}-{\vec k}_{2} &=&0,  \label{cons}
\end{eqnarray}%
where $E_{{\rm n}}\left( {\vec p}\right) $ and $E_{{\rm p}}\left( {\vec p}%
^{\prime }\right) $ are the in-medium energies of the neutron and the proton
respectively.

The energy exchange in the matter goes naturally on the temperature scale $%
\sim T$, which is small compared to typical kinetic energies of degenerate
particles. Therefore the momenta of in-medium fermions can be fixed at their
values at Fermi surfaces, which we denote as $p_{{\rm n}}$, $p_{{\rm p}}$
for the nucleons and $p_{l}$ for leptons respectively. Since the
antineutrino energy is $\omega _{1}\sim T$, and the antineutrino momentum $%
\left| {\vec k}_{1}\right| \sim T$ is much smaller than the momenta of other
particles, we can neglect the neutrino contributions in Eqs. (\ref{cons}).
Then the momentum conservation, ${\vec p}_{{\rm p}}+{\vec p}_{l}={\vec p}_{{\rm %
n}}$, implies the well known ''triangle'' condition, 
\begin{equation}
p_{{\rm p}}+p_{l}>p_{{\rm n}},  \label{triangle}
\end{equation}%
necessary for the Urca processes to operate. However, the momentum
conservation is necessary but insufficient condition for the direct beta
decay to occur. The energy conservation requires a one more condition%
\begin{equation}
E_{{\rm n}}\left( p_{{\rm n}}\right) -E_{{\rm p}}\left( p_{{\rm p}}\right) =%
\sqrt{\left( {\vec p}_{{\rm n}}-{\vec p}_{{\rm p}}\right) ^{2}+m_{l}^{2}},
\label{Econs}
\end{equation}%
Squaring of both sides of this equation gives%
\begin{equation}
\left( E_{{\rm n}}\left( p_{{\rm n}}\right) -E_{{\rm p}}\left( p_{{\rm p}%
}\right) \right) ^{2}-\left( {\vec p}_{{\rm n}}-{\vec p}_{{\rm p}}\right)
^{2}=m_{l}^{2}.  \label{Ec}
\end{equation}%
If we denote the energy-momentum transfer from the nucleon as 
\begin{equation}
K=\left( E_{{\rm n}}-E_{{\rm p}},{\vec p}_{{\rm n}}-{\vec p}_{{\rm p}}\right) ,
\label{K}
\end{equation}
the Eq. (\ref{Ec}) can be readily recognized as $K_{\mu }^{2}=m_{l}^{2}$.
This condition naturally arises from the time-like momentum of the final
lepton pair, $K=k_{1}+k_{2}\simeq k_{2}$, and can be satisfied only in the
presence of the energy gap between spectrums of protons and neutrons.

Consider now the condition of beta equilibrium, $\mu _{{\rm n}}-\mu _{{\rm p}%
}=\mu _{l}$, where $\mu _{{\rm n}}$, $\mu _{{\rm p}}$, and $\mu _{l}$ are
the chemical potentials of neutrons, protons, and leptons respectively. The
chemical potentials of degenerate particles can be approximated by their
individual Fermi energies, yielding 
\begin{equation}
E_{{\rm n}}\left( p_{{\rm n}}\right) -E_{{\rm p}}\left( p_{{\rm p}}\right) =%
\sqrt{p_{l}^{2}+m_{l}^{2}}  \label{chem}
\end{equation}%
Generally\footnote{%
We remind that, due to the modified Urca processes, the nuclear matter
attains chemical equilibrium even if the direct beta decay is forbidden. In
this case the Eq. (\ref{chem}) is valid but $p_{l}\neq \left| {\vec p}_{{\rm n%
}}-{\vec p}_{{\rm p}}\right| $.} this equation is not the same as Eq. (\ref%
{Econs}). However, if the ''triangle'' condition (\ref{triangle}) is
fulfilled then the direct neutron decay is open, and one has ${\vec p}_{l}=%
{\vec p}_{{\rm n}}-{\vec p}_{{\rm p}}$. In this case the Eq. (\ref{chem})
becomes identical to the energy consrvation (\ref{Econs}). Thus under beta
equilibrium, the ''triangle'' condition is the necessary and sufficient for
the direct neutron decay to occur. It should be emphasized however that the
stated conditions are consistent only in the presence of the energy gap
between protons and neutrons. Therefore the possibility of the direct
neutron decay depends substantially on the model of nuclear matter.

\subsection{Free gas model}

Consider, for example, a degenerate free gas consisting of neutrons,
protons, and electrons under beta equilibrium. In this case the energy gap
exists only due to the mass difference of a neutron and a proton. If we
denote the masses as $M_{{\rm n}}$ and $M_{{\rm p}}$ respectively, then the
corresponding Fermi energies are $E_{{\rm n}}\left( p_{{\rm n}}\right) =$ $%
\sqrt{M_{{\rm n}}^{2}+p_{{\rm n}}^{2}}$, and $E_{{\rm p}}\left( p_{{\rm p}%
}\right) =$ $\sqrt{M_{{\rm p}}^{2}+p_{{\rm p}}^{2}}$. Due to charge
neutrality the number density of electrons, $n_{e}\propto p_{l}^{3}$, equals
to the number density of protons, $n_{{\rm p}}\propto p_{{\rm p}}^{3}$. This
implies $p_{{\rm e}}=p_{{\rm p}}$ and the equation of chemical equilibrium (%
\ref{chem}) becomes of the form%
\begin{equation}
\sqrt{m_{{\rm e}}^{2}+p_{{\rm p}}^{2}}+\sqrt{M_{{\rm p}}^{2}+p_{{\rm p}}^{2}}%
=\sqrt{M_{{\rm n}}^{2}+p_{{\rm n}}^{2}}.  \label{consF}
\end{equation}%
Solution of this equation 
\begin{equation}
p_{{\rm p}}\left( p_{{\rm n}}\right) =\frac{1}{2\sqrt{M_{{\rm n}}^{2}+p_{%
{\rm n}}^{2}}}\sqrt{\left( p_{{\rm n}}^{2}+M_{{\rm n}}^{2}-\left( M_{{\rm p}%
}+m_{{\rm e}}\right) ^{2}\right) \left( p_{{\rm n}}^{2}+M_{{\rm n}%
}^{2}-\left( M_{{\rm p}}-m_{{\rm e}}\right) ^{2}\right) \allowbreak }
\label{pbeta}
\end{equation}%
$\allowbreak $ gives the proton Fermi momentum as a function of the neutron
Fermi momentum in the beta-equilibrated gas of protons and neutrons. As
required by the ''triangle'' condition, $2p_{{\rm p}}\geq p_{{\rm n}}$, the
direct neutron decay in such a medium is open when $p_{{\rm p}}$ is larger
than $p_{{\rm n}}/2$. So, solution of the equation%
\begin{equation}
p_{{\rm p}}\left( p_{{\rm n}}\right) =\frac{p_{{\rm n}}}{2}  \label{1}
\end{equation}%
$\allowbreak $gives the critical value of the neutron Fermi momentum above
which the direct neutron decay is forbidden. We find 
\begin{equation}
p_{{\rm n}}^{c}=\frac{\sqrt{\left( \left( M_{p}+M_{n}\right)
^{2}-m_{l}^{2}\right) \left( \left( M_{n}-M_{p}\right) ^{2}-m_{l}^{2}\right) 
}}{\sqrt{2M_{p}^{2}+2m_{l}^{2}-M_{n}^{2}}}=2.\,\allowbreak 381\,1\,MeV.
\label{2}
\end{equation}%
Thus the direct neutron decay is forbidden if the number density of neutrons
is larger than 
\begin{equation}
n_{n}^{c}=\frac{\left( p_{{\rm n}}^{c}\right) ^{3}}{3\pi ^{2}}=5.932\times
10^{31}\,cm^{-3}  \label{3}
\end{equation}%
This number density is much smaller than that typical for neutron star
cores, therefore the direct neutron decay in the cooling neutron stars can
occur only due to strong interactions.

Notice, some of models of strong interaction also forbid the direct neutron
decay. Consider, for example, a simple model for the baryon matter \cite%
{Wa74}, which contains fields for baryons and neutral scalar $\left( \sigma
\right) $ and vector $\left( \omega _{\mu }\right) $ mesons. This model
reproduces in a simple way the empirical NN scattering amplitude and the
bulk nuclear properties. However, the neutral scalar and vector mesons
equally interact with protons and neutrons, which therefore have identical
energy spectrums. In this case, the energy and momentum in the direct
neutron decay can not be conserved simultaneously.\ 

To allow the direct neutron decay, the model of nuclear matter must be
generalized to include some additional degrees of freedom and couplings,
which are able to create a large energy gap between possible energies of the
proton and neutron. For this purpose, besides isoscalar mesons $\sigma $ and 
$\omega $, the model should include also isovector mesons. The isovector
meson couples differently to protons and neutrons, thus creating the energy
gap necessary for the direct neutron decay.

\subsection{Field theoretical model. Mean field approximation.}

In the following we consider a self-consistent relativistic model of nuclear
matter in which baryons interact via exchange of isoscalar mesons $\sigma $
and $\omega $ and an isovector meson $\rho $ (See Appendix). In the mean
field approximation, when the contribution of mesons reduce to classical
condensate fields $\left\langle \sigma \right\rangle =\sigma _{0}$, $%
\left\langle \omega ^{\mu }\right\rangle =\omega _{0}\delta ^{\mu 0}$, $%
\left\langle {\vec b}^{\mu }\right\rangle \equiv \left( 0,0,\rho _{0}\right)
\delta ^{\mu 0}$, only the baryon fields must be quantized. This procedure
yields the following linear Dirac equation for the nucleon 
\begin{equation}
\left( i\partial _{\mu }\gamma ^{\mu }-g_{\omega }\gamma ^{0}\omega _{0}-%
\frac{1}{2}g_{\rho }\gamma ^{0}\rho _{0}\tau _{3}-\left( M-g_{\sigma }\sigma
_{0}\right) \right) \Psi \left( x\right) =0,  \label{DiracEq}
\end{equation}%
Here and below we denote as $\tau _{3}$, and $\tau _{\pm }=\left( \tau
_{1}\pm i\tau _{2}\right) /2$ the components of isospin operator, which act
on the isobaric doublet $\Psi \left( x\right) $ of nucleon field; $M=939\
MeV $ is the bare nucleon mass\footnote{%
As explained in previous Section, the mass difference of the proton and the
neutron can be neglected considering the nucleon as an isobaric doublet.}.

The stationary and uniform condensate fields equally shift the effective
masses 
\begin{equation}
M^{\ast }=M-g_{\sigma }\sigma _{0}  \label{Mef}
\end{equation}%
but lead to different potential energies of the proton and neutron 
\begin{equation}
U_{{\rm n}}=g_{\omega }\omega _{0}-\frac{1}{2}g_{\rho }\rho _{0},\,\ \ \ U_{%
{\rm p}}=g_{\omega }\omega _{0}+\frac{1}{2}g_{\rho }\rho _{0},  \label{Unp}
\end{equation}%
thus creating the energy gap $U_{{\rm n}}-U_{{\rm p}}=-g_{\rho }\rho _{0}$
between possible energies of protons and neutrons.

The exact solutions of Eq. (\ref{DiracEq}) can be found separately for
protons and for neutrons. In our case of a stationary, uniform system,
solution for the neutron is a spinor plane wave 
\begin{equation}
\psi _{{\rm n}}\left( x\right) =N_{{\rm n}}u_{{\rm n}}\exp \left( -iE_{{\rm n%
}}t+i{\vec {p} \vec {r}}\right) ,  \label{nvf}
\end{equation}%
where the neutron energy is given by $E_{{\rm n}}\left( {\vec p}\right) =%
\sqrt{{\vec p}^{2}+M^{\ast 2}}+U_{{\rm n}}$, and the spinor $u_{{\rm n}%
}\left( P\right) $ should be found from the following equation%
\begin{equation}
\left( \left( E_{{\rm n}}-U_{{\rm n}}\right) \gamma ^{0}-{\vec {\gamma}\vec {p}}%
-M^{\ast }\right) u_{{\rm n}}=0.  \label{unEq}
\end{equation}%
It is easily to understand that solution of this equation $u_{{\rm n}}=u_{%
{\rm n}}\left( P\right) $ is a free-like spinor constructed from the neutron 
{\em kinetic} momentum 
\begin{equation}
P^{\mu }=\left( E_{{\rm n}}-U_{{\rm n}},{\vec p}\right) =\left( \sqrt{{\vec p}%
^{2}+M^{\ast 2}},{\vec p}\right) ,  \label{kinN}
\end{equation}%
For the proton, one has%
\begin{equation}
\psi _{{\rm p}}\left( x\right) =N_{{\rm p}}u_{{\rm p}}\left( P^{\prime
}\right) \exp \left( -iE_{{\rm p}}t+i{\vec p}^{\prime }{\vec r}\right) ,
\label{pvf}
\end{equation}%
where the spinor $u_{{\rm p}}\left( P^{\prime }\right) $ obeys the equation%
\begin{equation}
\left( \left( E_{{\rm p}}-U_{{\rm p}}\right) \gamma ^{0}-{\vec {\gamma}\vec {p}}%
^{\prime }-M^{\ast }\right) u_{{\rm p}}=0  \label{upEq}
\end{equation}%
and depends respectively on the proton kinetic momentum 
\begin{equation}
P^{\prime \mu }=\left( E_{{\rm p}}-U_{{\rm p}},{\vec p}^{\prime }\right)
=\left( \sqrt{{\vec p}^{\prime 2}+M^{\ast 2}},{\vec p}^{\prime }\right) .
\label{kinP}
\end{equation}%
In what follows we denote by $\varepsilon =\sqrt{{\vec p}^{2}+M^{\ast 2}},\,\
\varepsilon ^{\prime }=\sqrt{{\vec p}^{\prime 2}+M^{\ast 2}}$ the kinetic
energy of the neutron and the proton respectively. So the normalization
factors are of the form 
\begin{equation}
N_{{\rm n}}=\frac{1}{\sqrt{2\varepsilon }},\,\ \ \ \ \ N_{{\rm p}}=\frac{1}{%
\sqrt{2\varepsilon ^{\prime }}},  \label{norm}
\end{equation}%
and the single-particle energies are $E_{{\rm n}}\left( {\vec p}\right)
=\varepsilon +U_{{\rm n}}$, and $E_{{\rm p}}\left( {\vec p}^{\prime }\right)
=\varepsilon ^{\prime }+U_{{\rm p}}$.

At the Fermi surfaces the single-particle energies are $E_{{\rm n}}\left( p_{%
{\rm n}}\right) =\sqrt{M^{\ast 2}+p_{{\rm n}}^{2}}+U_{{\rm n}}$, and $E_{%
{\rm p}}\left( p_{{\rm p}}\right) =\sqrt{M^{\ast 2}+p_{{\rm p}}^{2}}+U_{{\rm %
p}}$. In this case the Eq. (\ref{Ec}) takes the form%
\begin{equation}
\left( \sqrt{M_{{\rm n}}^{\ast 2}+p_{{\rm n}}^{2}}-\sqrt{M_{{\rm p}}^{\ast
2}+p_{{\rm p}}^{2}}+U_{{\rm n}}-U_{{\rm p}}\right) ^{2}-\left( {\vec p}_{{\rm %
n}}-{\vec p}_{{\rm p}}\right) ^{2}=m_{l}^{2}.  \label{EcU}
\end{equation}%
Typically $U_{{\rm n}}-U_{{\rm p}}=-g_{\rho }\rho _{0}\sim 100\,MeV$, so
this equation can be readily satisfied simultaneously with the momentum
conservation ${\vec p}_{{\rm n}}-{\vec p}_{{\rm p}}={\vec p}_{l}$.

\subsection{Nonconservation of the charged vector current of nucleons.}

The Lagrangian density (\ref{Lagr}) ensures a conserved isovector current %
\cite{SW97} 
\begin{equation}
{\vec T}^{\mu }=\frac{1}{2}\bar{\psi}\gamma ^{\mu }{\vec \tau }\psi +{\vec b}%
_{\nu }\times {\vec B}^{\nu \mu },\,\ \ \ \ \ \partial _{\mu }{\vec T}^{\mu
}=0,  \label{CIC}
\end{equation}%
where ${\vec B}^{\mu \nu }=\partial ^{\mu }{\vec b}^{\nu }-\partial ^{\nu }%
{\vec b}^{\mu }$ is the $\rho $ meson field strength tensor. Besides the
directly nucleon contribution the conserved current (\ref{CIC}) includes
also the contribution of the isovector field ${\vec b}^{\mu }$, which obeys
the field equations 
\begin{equation}
\partial _{\nu }{\vec B}^{\nu \mu }+m_{\rho }^{2}{\vec b}^{\mu }=\frac{1}{2}%
g_{\rho }\bar{\psi}\gamma ^{\mu }{\vec \tau }\psi ,\,\ \ \ \ \partial ^{\nu }%
{\vec b}_{\nu }=0.  \label{bEq}
\end{equation}%
By the use of Eq. (\ref{bEq}) the condition $\partial _{\mu }{\vec T}^{\mu
}=0 $ may be transformed as%
\begin{equation}
i\partial _{\mu }\left( \bar{\psi}\gamma ^{\mu }{\vec \tau }\psi \right)
=g_{\rho }{\vec b}_{\mu }\times \bar{\psi}\gamma ^{\mu }{\vec \tau }\psi .
\label{MFcic}
\end{equation}%
In the mean field approximation this gives 
\begin{equation}
i\partial _{\mu }\left( \bar{\psi}\gamma ^{\mu }\tau _{+}\psi \right)
=-g_{\rho }\rho _{0}\bar{\psi}\gamma ^{0}\tau _{+}\psi .  \label{ncnc}
\end{equation}%
By introducing the covariant derivative 
\begin{equation}
D_{\mu }=\left( \frac{\partial }{\partial t}-ig_{\rho }\rho _{0},\,{\vec %
\nabla }\right) ,  \label{cd}
\end{equation}%
we can recast the Eq. (\ref{ncnc}) to the following form%
\begin{equation}
D_{\mu }\left( \bar{\psi}\gamma ^{\mu }\tau _{+}\psi \right) =0.  \label{cc}
\end{equation}%
At the level of matrix elements this can be written as%
\begin{equation}
\bar{u}_{{\rm p}}\left( P^{\prime }\right) q_{\mu }\gamma ^{\mu }u_{{\rm n}%
}\left( P\right) =0,  \label{orthq}
\end{equation}%
where $q_{\mu }$ is the kinetic momentum transfer 
\begin{equation}
q^{\mu }=\left( E_{{\rm n}}-E_{{\rm p}}+g_{\rho }\rho _{0},\,{\vec {p}-\vec {p}^{\prime}}
\right) =\left( \varepsilon -\varepsilon ^{\prime },{\vec {p}-\vec {p}^{\prime}}
\right)  \label{q}
\end{equation}%
Thus the matrix element of the transition current is orthogonal to the {\em %
kinetic} momentum transfer but not to the total momentum transfer from the
nucleon\footnote{%
Some consequences of this fact for the weak response functions of the medium
are discussed in \cite{PLB}}. Note that this effect originates not from a
special form (\ref{CIC}) of the conserved isovector current in the medium
but is caused by the energy gap between the proton and neutron spectrums.
This follows directly from the Dirac equation (\ref{DiracEq}), which ensures
the Eq. (\ref{orthq}) with 
\begin{equation}
q^{\mu }=P^{\mu }-P^{\prime \mu }=\left( E_{{\rm n}}-E_{{\rm p}}-U_{{\rm n}%
}+U_{{\rm p}},\,{\vec {p}-\vec {p}^{\prime}}\right) ,  \label{qq}
\end{equation}%
which coincides with Eq. (\ref{q}) because $U_{{\rm n}}-U_{{\rm p}}=-g_{\rho
}\rho _{0}$.

\section{Matrix element of the neutron beta decay.}

In the lowest order in the Fermi weak coupling constant $G_{F}$, the matrix
element of the neutron beta decay is of the form%
\begin{eqnarray}
\left\langle f\right| \left( S-1\right) \left| i\right\rangle &=&-i\frac{%
G_{F}C}{\sqrt{2}}N_{{\rm n}}N_{{\rm p}}\bar{u}_{l}\left( k_{2}\right) \gamma
_{\mu }\left( 1+\gamma _{5}\right) \nu \left( -k_{1}\right) \,_{{\rm p}%
}\left\langle P^{\prime }\right| J^{\mu }\left( 0\right) \left|
P\right\rangle _{{\rm n}}\times  \nonumber \\
&&\times \left( 2\pi \right) ^{4}\delta \left( E_{{\rm n}}-E_{{\rm p}%
}-\omega _{1}-\omega _{2}\right) \delta \left( {\vec p}-{\vec p}^{\prime }-%
{\vec k}_{1}-{\vec k}_{2}\right) ,  \label{S}
\end{eqnarray}%
were $C=\cos \theta _{C}=0\allowbreak .\,\allowbreak 973$ is the Cabibbo
factor. The effective charged weak current in the medium consists of the
polar vector and the axial vector, $J^{\mu }\left( x\right) =V^{\mu }\left(
x\right) +A^{\mu }\left( x\right) $.

Our goal now is to derive the nucleon matrix element of the charged weak
current in the medium. Consider first the polar-vector contribution. By the
use of the isovector current (\ref{CIC}) one can construct the conserved
electromagnetic current in the medium%
\begin{equation}
J_{{\rm em}}^{\mu }=\frac{1}{2}\bar{\psi}\gamma ^{\mu }\psi +T_{3}^{\mu }+%
\frac{1}{2M}\,\partial _{\nu }\left( \bar{\Psi}\lambda \sigma ^{\mu \nu
}\Psi \right) ,\,\ \ \ \ \ \ \partial _{\mu }J_{{\rm em}}^{\mu }=0.
\label{Jem}
\end{equation}%
The last term in\ Eq. (\ref{Jem}) is the Pauli contribution, where $2\sigma
^{\mu \nu }=\gamma ^{\mu }\gamma ^{\nu }-\gamma ^{\nu }\gamma ^{\mu }$ and 
\begin{equation}
\lambda =\lambda _{{\rm p}}\frac{1}{2}\left( 1+\tau _{3}\right) +\lambda _{%
{\rm n}}\frac{1}{2}\left( 1-\tau _{3}\right) .  \label{lam}
\end{equation}%
In the mean field approximation, we replace the magnetic formfactors of the
nucleon with anomalous magnetic moments of the proton and the neutron, $%
\lambda _{{\rm p}}=1.7928$ and $\lambda _{{\rm n}}=-1.9132$.

By the conserved-vector-current theory (CVC), the nucleon matrix element of
the charged vector weak current is given by 
\begin{equation}
_{{\rm p}}{}\left\langle P^{\prime }\right| V^{\mu }\left| P\right\rangle _{%
{\rm n}}=\,_{{\rm p{}}}\left\langle P^{\prime }\right| J_{{\rm em}}^{\mu
}\left| P\right\rangle _{{\rm p}}-\,_{{\rm n}}{}\left\langle P^{\prime
}\right| J_{{\rm em}}^{\mu }\left| P\right\rangle _{{\rm n}}.  \label{VJ}
\end{equation}%
For the electromagnetic transitions one has%
\begin{equation}
_{{\rm p}}{}\left\langle P^{\prime }\right| J_{{\rm em}}^{\mu }\left(
0\right) \left| P\right\rangle _{{\rm p}}=\,\bar{u}_{{\rm p}}\left(
P^{\prime }\right) \left( \gamma ^{\mu }+\frac{1}{2M}\lambda _{{\rm p}%
}\,\sigma ^{\mu \nu }q_{\nu }\right) u_{{\rm p}}\left( P\right) ,  \label{Jp}
\end{equation}%
\begin{equation}
_{{\rm n}}{}\left\langle P^{\prime }\right| J_{{\rm em}}^{\mu }\left(
0\right) \left| P\right\rangle _{{\rm n}}=\,\bar{u}_{{\rm n}}\left(
P^{\prime }\right) \left( \frac{1}{2M}\lambda _{{\rm n}}\,\sigma ^{\mu \nu
}q_{\nu }\right) u_{{\rm n}}\left( P\right)  \label{Jn}
\end{equation}%
with $q_{\mu }=P-P^{\prime }$, as given by Eq. (\ref{q}). Thus, in the mean
field approximation, we obtain 
\begin{equation}
_{{\rm p}}{}\!\left\langle P^{\prime }\right| V^{\mu }\left( 0\right) \left|
P\right\rangle _{{\rm n}}=\bar{u}_{{\rm p}}\left( P^{\prime }\right) \left[
\gamma ^{\mu }+\frac{\lambda _{{\rm p}}-\lambda _{{\rm n}}}{2M}\sigma ^{\mu
\nu }q_{\nu }\right] u_{{\rm n}}\left( P\right) .  \label{wcv}
\end{equation}%
The second term in Eq. (\ref{wcv}), describes the weak magnetism effects. By
the use of the Dirac Eqs. (\ref{unEq}) and (\ref{upEq}) for the nucleon
spinors we find 
\begin{equation}
_{{\rm p}}{}\!\left\langle P^{\prime }\right| q_{\mu }V^{\mu }\left(
0\right) \left| P\right\rangle _{{\rm n}}=0  \label{qV}
\end{equation}%
in accord with Eq. (\ref{orthq}).

Consider now the axial-vector charged current. This current is responsible
for both the $np$ transitions and the pion decay. In the limit of chiral
symmetry, $m_{\pi }\rightarrow 0$, the axial-vector current must be
conserved. In the medium with $\rho $ meson condensate, this implies 
\begin{equation}
\lim_{m_{\pi }\rightarrow 0}D_{\mu }A^{\mu }\left( x\right) =0,  \label{DA}
\end{equation}%
where the covariant derivative $D_{\mu }$ is defined by Eq. (\ref{cd}). At
the finite mass of a pion, $m_{\pi }$, the axial-vector charged current is
connected to the field $\pi _{-}\left( x\right) =\left( \pi _{1}+i\pi
_{2}\right) /\sqrt{2}$ of $\pi ^{-}$ meson. For a free space, this relation
is known as the hypothesis of partial conservation of the axial current
(PCAC). In the medium the PCAC takes the form 
\begin{equation}
D_{\mu }A^{\mu }\left( x\right) =m_{\pi }^{2}f_{\pi }\pi _{-}\left( x\right)
,  \label{PCAC}
\end{equation}%
where $m_{\pi }=139\,MeV$ is the mass of $\pi $-meson, and $f_{\pi }$ is the
pion decay constant.

With allowing for interactions of the pions with nucleons and $\rho $ mesons
the Lagrangian density for the pion field is of the form \cite{SW97}%
\begin{equation}
{\cal L}_{\pi }=\frac{1}{2}\left[ \left( \partial _{\mu }{\vec \pi }-g_{\rho }%
{\vec b}_{\mu }\times {\vec \pi }\right) \cdot \left( \partial ^{\mu }{\vec \pi 
}-g_{\rho }{\vec b}^{\mu }\times {\vec \pi }\right) -m_{\pi }^{2}{\vec \pi
\cdot \vec \pi }\right] +ig_{\pi }\bar{\psi}\gamma _{5}{\vec \tau \cdot \vec \pi }\psi 
\text{.}  \label{Lpi}
\end{equation}%
In the mean field approximation this results in the following equation for
the field of $\pi ^{-}$ meson%
\begin{equation}
\left( \left( i\partial ^{0}+g_{\rho }\rho _{0}\right) ^{2}-\left( i\nabla
\right) ^{2}-m_{\pi }^{2}\right) \pi _{-}\left( x\right) =-\sqrt{2}ig_{\pi }%
\bar{\psi}_{{\rm p}}\gamma _{5}\psi _{{\rm n}},  \label{pifield}
\end{equation}%
where $g_{\pi }$ is the pion-nucleon coupling constant.

For the nucleon transition of our interest, the Eq. (\ref{PCAC}) gives 
\begin{equation}
\,_{{\rm p}}\!\left\langle P^{\prime }\right| q_{\mu }A^{\mu }\left(
0\right) \left| P\right\rangle _{{\rm n}}=im_{\pi }^{2}f_{\pi }\,_{{\rm p}%
}\left\langle P^{\prime }\right| \pi _{-}\left( 0\right) \left|
P\right\rangle _{{\rm n}}  \label{npPCAC}
\end{equation}%
Here the right-hand side can be calculated by the use of Eq. (\ref{pifield}%
). We obtain%
\begin{equation}
\,_{{\rm p}}\!\left\langle P^{\prime }\right| q_{\mu }A^{\mu }\left(
0\right) \left| P\right\rangle _{{\rm n}}=-\frac{\sqrt{2}m_{\pi }^{2}f_{\pi
}g_{\pi }}{m_{\pi }^{2}-q^{2}}\bar{u}_{{\rm p}}\left( P^{\prime }\right)
\gamma _{5}u_{{\rm n}}\left( P\right) \text{.}  \label{mePCAC}
\end{equation}%
This equation allows to derive the nucleon matrix element of the
axial-vector charged current. Really, to construct the axial-vector matrix
element of the charged current, caused by the nucleon transition, we have
only two independent pseudovectors, consistent with invariance of strong
interactions under $T_{2}$ isospin transformation, namely: $\bar{u}_{{\rm p}%
}\left( P^{\prime }\right) \gamma ^{\mu }\gamma _{5}u_{{\rm n}}\left(
P\right) $, and $\bar{u}_{{\rm p}}\left( P^{\prime }\right) q^{\mu }\gamma
_{5}u_{{\rm n}}\left( P\right) $. This means that the matrix element of the
axial-vector charged current is of following general form 
\begin{equation}
_{{\rm p}}\!\left\langle P^{\prime }\right| A^{\mu }\left( 0\right) \left|
P\right\rangle _{{\rm n}}=C_{A}\,\bar{u}_{{\rm p}}\left( P^{\prime }\right)
\left( \gamma ^{\mu }\gamma _{5}+F_{q}\,q^{\mu }\gamma _{5}\right) u_{{\rm n}%
}\left( P\right) .  \label{AmeG}
\end{equation}%
Here, in the mean field approximation, we set $C_{A}\simeq 1.26$, while $%
F_{q}$ is the form-factor to be chosen to satisfy the Eq. (\ref{mePCAC}),
which now reads%
\begin{equation}
C_{A}\,\left( -2M^{\ast }+F_{q}\,q^{2}\right) \bar{u}_{{\rm p}}\left(
P^{\prime }\right) \gamma _{5}u_{{\rm n}}\left( P\right) =-\frac{\sqrt{2}%
m_{\pi }^{2}f_{\pi }g_{\pi }}{m_{\pi }^{2}-q^{2}}\bar{u}_{{\rm p}}\left(
P^{\prime }\right) \gamma _{5}u_{{\rm n}}\left( P\right) .  \label{lhs}
\end{equation}%
To obtain this we used the Dirac Eqs. (\ref{unEq}) and (\ref{upEq}) for the
nucleon spinors. Thus%
\begin{equation}
C_{A}\,\left( 2M^{\ast }-F_{q}\,q^{2}\right) =\frac{\sqrt{2}m_{\pi
}^{2}f_{\pi }g_{\pi }}{m_{\pi }^{2}-q^{2}}.  \label{Feq}
\end{equation}%
In the mean field approximation, we assume that the coupling constants are
independent of the momentum transfer. By setting $q^{2}=0$ in Eq. (\ref{Feq}%
) we obtain the Goldberger - Treiman relation%
\begin{equation}
f_{\pi }g_{\pi }=\sqrt{2}M^{\ast }C_{A}.  \label{GT}
\end{equation}%
By inserting this in (\ref{Feq}) we find%
\begin{equation}
F_{q}\,=-\,\frac{2M^{\ast }}{\left( m_{\pi }^{2}-q^{2}\right) }.  \label{Fq}
\end{equation}%
Thus, with taking into account Eqs. (\ref{wcv}), (\ref{AmeG}), and (\ref{Fq}%
), the total matrix element of the neutron beta decay is found to be%
\begin{eqnarray}
{\cal M}_{fi} &=&-i\frac{G_{F}C}{\sqrt{2}}\bar{u}_{l}\left( k_{2}\right)
\gamma _{\mu }\left( 1+\gamma _{5}\right) \nu \left( -k_{1}\right) \,\times
\label{mend} \\
&&\times \bar{u}_{{\rm p}}\left( P^{\prime }\right) \left[ C_{V}\gamma ^{\mu
}+\frac{1}{2M}C_{M}\sigma ^{\mu \nu }q_{\nu }+C_{A}\left( \gamma ^{\mu
}\gamma _{5}+F_{q}\,q^{\mu }\gamma _{5}\right) \right] u_{{\rm n}}\left(
P\right) ,  \nonumber
\end{eqnarray}%
where, in the mean field approximation, we assume 
\begin{equation}
C_{V}=1,\,\ \ \ \ \ C_{M}=\lambda _{p}-\lambda _{n}\simeq 3.7,\,\ \ \
C_{A}=1.26.  \label{ff}
\end{equation}%
Note that the matrix element obtained is of the same form as that for the
neutron decay in a free space, but with the total momentum transfer replaced
with the kinetic momentum transfer. Due to the difference in the neutron and
proton potential energy, the kinetic momentum transfer 
\begin{equation}
q=P-P^{\prime }=\left( \varepsilon -\varepsilon ^{\prime },{\vec p-\vec p^\prime
}\right)  \label{kintr}
\end{equation}%
to be used in the matrix element (\ref{mend}) differs from the total
momentum of the final lepton pair 
\begin{equation}
K=\left( \varepsilon -\varepsilon ^{\prime }+U_{{\rm n}}-U_{{\rm p}},\,{\vec %
p-\vec p^\prime }\right)  \label{Q}
\end{equation}%
This ensures $K^{2}>0$, while $q^{2}=\left( \varepsilon -\varepsilon
^{\prime }\right) ^{2}-\left( {\vec p}-{\vec p}^{\prime }\right) ^{2}<0$ .

The square of the matrix element of the reaction summed over spins of
initial and final particles is found to be\footnote{%
Here we rectify an error made in the journal version of the paper. The
correct expression can be obtained from that published in \cite{PLB02}, \cite%
{NPA02} by simple replacement $C_{M}\rightarrow C_{M}/2$. The author is
grateful to M. Prakash and S. Ratkovi\'{c} who have pointed out this error.}%
: 
\begin{eqnarray}
\left| {\cal M}_{fi}\right| ^{2} &=&32G_{F}^{2}C^{2}\left[ \left(
C_{A}^{2}-C_{V}^{2}\right) M^{\ast 2}\left( k_{1}k_{2}\right) +\left(
C_{A}-C_{V}\right) ^{2}\left( k_{1}P_{2}\right) \left( k_{2}P_{1}\right)
\right.  \nonumber \\
&&+\left( C_{A}+C_{V}\right) ^{2}\left( k_{1}P_{1}\right) \left(
k_{2}P_{2}\right)  \nonumber \\
&&+C_{M}\frac{M^{\ast }}{M}\left[ 2C_{A}\left( \left( k_{1}P_{1}\right)
\left( k_{2}P_{2}\right) -\left( k_{1}P_{2}\right) \left( k_{2}P_{1}\right)
\right) \right.  \nonumber \\
&&\left. +C_{V}\left( \left( k_{1}k_{2}\right) \left( P_{1}P_{2}-M^{\ast
2}\right) -\left( k_{1}P_{1}-k_{1}P_{2}\right) \left(
k_{2}P_{1}-k_{2}P_{2}\right) \right) \right]  \nonumber \\
&&-\frac{C_{M}^{2}}{4M^{2}}\left[ M^{\ast 2}\left( k_{1}P_{2}\right) \left(
3\left( k_{2}P_{2}\right) -\left( k_{2}P_{1}\right) \right) \right. 
\nonumber \\
&&+M^{\ast 2}\left( k_{1}P_{1}\right) \left( 3\left( k_{2}P_{1}\right)
-\left( k_{2}P_{2}\right) \right) +\left( k_{1}k_{2}\right) \left(
P_{1}P_{2}-M^{\ast 2}\right) ^{2}  \nonumber \\
&&\left. -\left( k_{1}P_{1}+k_{1}P_{2}\right) \left(
k_{2}P_{1}+k_{2}P_{2}\right) \left( P_{1}P_{2}\right) \right]  \nonumber \\
&&+C_{A}^{2}F_{q}\left( 2M^{\ast }+F_{q}\left( M^{\ast 2}-\left(
P_{1}P_{2}\right) \right) \right) \left[ \left( k_{1}k_{2}\right) \left(
M^{\ast 2}-\left( P_{1}P_{2}\right) \right) \right.  \nonumber \\
&&\left. \left. -\left( k_{1}P_{1}-k_{1}P_{2}\right) \left(
k_{2}P_{1}-k_{2}P_{2}\right) \right] \right]  \label{MatrEl}
\end{eqnarray}%
with $P_{1}=\left( \varepsilon ,{\vec p}\right) $ and $P_{2}=\left(
\varepsilon ^{\prime },{\vec p}^{\prime }\right) $.

\section{Neutrino energy losses}

We consider the total energy which is emitted into neutrino and antineutrino
per unit volume and time. Within beta equilibrium, the inverse reaction $%
p+l\rightarrow n+\nu _{l}$ corresponding to a capture of the lepton $l$,
gives the same emissivity as the beta decay, but in neutrinos. Thus, the
total energy loss $Q$ for the Urca processes is twice more than that caused
by the beta decay. Taking this into account by Fermi's ''golden'' rule we
have%
\begin{eqnarray}
Q &=&2\int \frac{d^{3}k_{2}d^{3}k_{1}d^{3}pd^{3}p^{\prime }}{(2\pi
)^{12}2\omega _{2}2\omega _{1}2\varepsilon 2\varepsilon ^{\prime }}\left| 
{\cal M}_{fi}\right| ^{2}\omega _{1}\,f_{{\rm n}}\left( 1-f_{{\rm p}}\right)
\left( 1-f_{l}\right)  \nonumber \\
&&\times \left( 2\pi \right) ^{4}\delta \left( E_{{\rm n}}\left( {\vec p}%
\right) -E_{{\rm p}}\left( {\vec p}^{\prime }\right) -\omega _{1}-\omega
_{2}\right) \delta \left( {\vec p}-{\vec p}^{\prime }-{\vec k}_{1}-{\vec k}%
_{2}\right) .  \label{gold}
\end{eqnarray}%
Antineutrinos are assumed to be freely escaping. The distribution function
of initial neutrons as well as blocking of final states of the proton and
the lepton $l$ are taken into account by the Pauli blocking-factor $\,f_{%
{\rm n}}\left( 1-f_{{\rm p}}\right) \left( 1-f_{l}\right) $. The Fermi-Dirac
distribution function of leptons is given by 
\begin{equation}
f_{l}\left( \omega _{2}\right) =\frac{1}{\exp \left( \omega _{2}-\mu
_{l}\right) /T+1},  \label{fl}
\end{equation}%
while the individual Fermi distributions of nucleons are of the form 
\begin{equation}
f_{{\rm n}}\left( \varepsilon \right) =\frac{1}{\exp \left( \left(
\varepsilon +U_{{\rm n}}-\mu _{{\rm n}}\right) /T\right) +1},  \label{fn}
\end{equation}%
\begin{equation}
f_{{\rm p}}\left( \varepsilon ^{\prime }\right) =\frac{1}{\exp \left( \left(
\varepsilon ^{\prime }+U_{{\rm p}}-\mu _{{\rm p}}\right) /T\right) +1},
\label{fp}
\end{equation}

By neglecting the chemical potential of escaping neutrinos, we can write the
condition of chemical equilibrium as $\mu _{l}=\mu _{{\rm n}}-\mu _{{\rm p}}$%
. Then by the use of the energy conservation equation, $\varepsilon +U_{{\rm %
n}}=\varepsilon ^{\prime }+U_{{\rm p}}+\omega _{2}+\omega _{1}$, and taking
the total energy of the final lepton and antineutrino as $\omega _{2}+\omega
_{1}=\mu _{l}+\omega ^{\prime }$ we can recast the blocking-factor as 
\begin{eqnarray}
&&\,f_{{\rm n}}\left( \varepsilon \right) \left( 1-f_{{\rm p}}\left(
\varepsilon ^{\prime }\right) \right) \left( 1-f_{l}\left( \omega
_{2}\right) \right)  \nonumber \\
&\equiv &f_{{\rm n}}\left( \varepsilon \right) \left( 1-f_{{\rm n}}\left(
\varepsilon -\omega ^{\prime }\right) \right) \left( 1-f_{l}\left( \mu
_{l}+\omega ^{\prime }-\omega _{1}\right) \right) ,  \label{block}
\end{eqnarray}%
where $\omega ^{\prime }\sim T$.

Furthermore, since the antineutrino energy is $\omega _{1}\sim T$, and the
antineutrino momentum $\left| {\vec k}_{1}\right| \sim T$ is much smaller
than the momenta of other particles, we can neglect the neutrino
contributions in the energy-momentum conserving delta-functions%
\begin{eqnarray}
&&\delta \left( \varepsilon +U_{{\rm n}}-\varepsilon ^{\prime }-U_{{\rm p}%
}-\omega _{1}-\omega _{2}\right) \delta \left( {\vec p}-{\vec p}^{\prime }-%
{\vec k}_{1}-{\vec k}_{2}\right)  \nonumber \\
&\simeq &\delta \left( \varepsilon +U_{{\rm n}}-\varepsilon ^{\prime }-U_{%
{\rm p}}-\omega _{2}\right) \delta \left( {\vec p}-{\vec p}^{\prime }-{\vec k}%
_{2}\right)  \label{delE}
\end{eqnarray}%
and perform integral over $d^{3}p^{\prime }$ to obtain ${\vec p}^{\prime }=%
{\vec p}-{\vec k}_{2}$ in the next integrals.

Nucleons and leptons in the neutron star core are strongly degenerate,
therefore the main contribution to the integral (\ref{gold}) comes from
narrow regions near the corresponding Fermi momenta. Thus we can set $\left| 
{\vec p}\right| =p_{{\rm n}}$, $\left| {\vec k}_{2}\right| =p_{l}$ in all
smooth functions under the integral.

The energy of the final lepton is close to its Fermi energy $\mu _{l}=\mu _{%
{\rm n}}-\mu _{{\rm p}}$. Here the chemical potentials of nucleons can be
approximated by their individual Fermi energies $\mu _{{\rm n}}=\varepsilon
_{{\rm n}}+U_{{\rm n}}$, $\mu _{{\rm p}}=\varepsilon _{{\rm p}}+U_{{\rm p}}$%
. This allows us to transform the energy-conserving $\delta $-function as 
\begin{eqnarray}
&&\delta (\varepsilon _{{\rm n}}-\sqrt{p_{{\rm n}}^{2}+p_{l}^{2}-2p_{{\rm n}%
}p_{l}\cos \theta _{l}+M^{\ast \,2}}+U_{{\rm n}}-U_{{\rm p}}-\mu _{l}) 
\nonumber \\
&=&\frac{\varepsilon _{{\rm p}}}{p_{{\rm n}}p_{l}}\delta \left( \cos \theta
_{l}-\frac{1}{2p_{{\rm n}}p_{l}}\left( p_{{\rm n}}^{2}-p_{{\rm p}%
}^{2}+p_{l}^{2}\right) \right) ,  \label{delfun}
\end{eqnarray}%
where $\theta _{l}$ is the angle between the momentum ${\vec p}_{{\rm n}}$ of
the initial neutron and the momentum ${\vec p}_{l}$ of the final lepton.
Notice, when the baryon and lepton momenta are at their individual Fermi
surfaces, the $\delta $- function (\ref{delfun}) does not vanish only if $p_{%
{\rm p}}+p_{l}>p_{{\rm n}}$.

Further we use the particular frame with $Z$-axis directed along the neutron
momentum ${\vec p}_{{\rm n}}$. Then%
\begin{eqnarray}
P_{1} &=&(0,\,0,\,p_{{\rm n}},\,\varepsilon _{{\rm n}})  \nonumber \\
k_{1} &=&\omega _{1}\left( \sin \theta _{\nu },\,0,\,\cos \theta _{\nu
},\,1\right)  \nonumber \\
k_{2} &=&\left( p_{l}\sin \theta _{l}\cos \varphi _{l},\,p_{l}\sin \theta
_{l}\sin \varphi _{l},\,p_{l}\cos \theta _{\nu },\,\mu _{l}\right)
\label{Pk}
\end{eqnarray}%
The energy-momentum of the final proton is defined by conservation laws:%
\begin{equation}
P_{2}=\left( -p_{l}\sin \theta _{l}\cos \varphi _{l},\,-p_{l}\sin \theta
_{l}\sin \varphi _{l},\,p_{{\rm n}}-p_{l}\cos \theta _{\nu },\,\varepsilon _{%
{\rm p}}\right)  \label{P2}
\end{equation}%
Insertion of (\ref{Pk}) and (\ref{P2}) in the square of the matrix element (%
\ref{MatrEl}) yields a rather cumbersome expression, which, however, is
readily integrable over solid angles of the particles.

Since we focus on the actually important case of degenerate nucleons and
leptons, we may consider the neutrino energy losses to the lowest accuracy
in $T/\mu _{l}$. Then the remaining integration reduces to the factor 
\begin{eqnarray}
&&\int d\omega _{1}\omega _{1}^{3}d\omega ^{\prime }d\varepsilon \,f_{{\rm n}%
}\left( \varepsilon \right) \left( 1-f_{{\rm n}}\left( \varepsilon -\omega
^{\prime }\right) \right) \left( 1-f_{l}\left( \mu _{l}+\omega ^{\prime
}-\omega _{1}\right) \right)  \nonumber \\
&\simeq &\int_{-\infty }^{\infty }d\omega ^{\prime }\frac{\omega ^{\prime }}{%
\exp \omega ^{\prime }/T-1}\int_{0}^{\infty }d\omega _{1}\frac{\omega
_{1}^{3}}{1+\exp \left( \omega _{1}-\omega ^{\prime }\right) /T}=\frac{457}{%
5040}\pi ^{6}T^{6}.  \label{int}
\end{eqnarray}%
Finally the neutrino emissivity is found to be of the form\footnote{%
See footnote to Eq. (\ref{MatrEl})}:%
\begin{eqnarray}
Q &=&\,\frac{457\pi }{10\,080}G_{F}^{2}C^{2}T^{6}\Theta \left( p_{l}+p_{{\rm %
p}}-p_{{\rm n}}\right) \left\{ \left( C_{A}^{2}-C_{V}^{2}\right) M^{\ast
2}\mu _{l}\right.  \nonumber \\
&&+\frac{1}{2}\left( C_{V}^{2}+C_{A}^{2}\right) \left[ 4\varepsilon _{{\rm n}%
}\varepsilon _{{\rm p}}\mu _{l}-\left( \varepsilon _{{\rm n}}-\varepsilon _{%
{\rm p}}\right) \left( \left( \varepsilon _{{\rm n}}+\varepsilon _{{\rm p}%
}\right) ^{2}-p_{l}^{2}\right) \right]  \nonumber \\
&&+C_{V}C_{M}\frac{M^{\ast }}{2M}\left[ 2\left( \varepsilon _{{\rm n}%
}-\varepsilon _{{\rm p}}\right) p_{l}^{2}-\left( 3\left( \varepsilon _{{\rm n%
}}-\varepsilon _{{\rm p}}\right) ^{2}-p_{l}^{2}\right) \mu _{l}\right] 
\nonumber \\
&&+C_{A}\left( C_{V}+\frac{M^{\ast }}{M}C_{M}\right) \left( \varepsilon _{%
{\rm n}}+\varepsilon _{{\rm p}}\right) \left( p_{l}^{2}-\left( \varepsilon _{%
{\rm n}}-\varepsilon _{{\rm p}}\right) ^{2}\right)  \nonumber \\
&&+C_{M}^{2}\frac{1}{16M^{2}}\left[ 8M^{\ast 2}\left( \varepsilon _{{\rm n}%
}-\varepsilon _{{\rm p}}\right) \left( p_{l}^{2}-\left( \varepsilon _{{\rm n}%
}-\varepsilon _{{\rm p}}\right) \mu _{l}\right) \right.  \nonumber \\
&&+\left( p_{l}^{2}-\left( \varepsilon _{{\rm n}}-\varepsilon _{{\rm p}%
}\right) ^{2}\right) \left( 2\varepsilon _{{\rm n}}^{2}+2\varepsilon _{{\rm p%
}}^{2}-p_{l}^{2}\right) \mu _{l}  \nonumber \\
&&\left. -\left( p_{l}^{2}-\left( \varepsilon _{{\rm n}}-\varepsilon _{{\rm p%
}}\right) ^{2}\right) \left( \varepsilon _{{\rm n}}+\varepsilon _{{\rm p}%
}\right) ^{2}\left( 2\varepsilon _{{\rm n}}-2\varepsilon _{{\rm p}}-\mu
_{l}\right) \right]  \nonumber \\
&&\left. -C_{A}^{2}M^{\ast 2}\Phi \left( 1+m_{\pi }^{2}\Phi \right) \left[
\mu _{l}\left( \left( \varepsilon _{{\rm n}}-\varepsilon _{{\rm p}}\right)
^{2}+p_{l}^{2}\right) -2\left( \varepsilon _{{\rm n}}-\varepsilon _{{\rm p}%
}\right) p_{l}^{2}\right] \right\}  \label{QMFA}
\end{eqnarray}

with $\Theta \left( x\right) =1$ if $x\geq 0$ and zero otherwise. In the
above, the last term, with $\allowbreak $ 
\begin{equation}
\Phi =\frac{1}{m_{\pi }^{2}+p_{l}^{2}-\left( \varepsilon _{{\rm n}%
}-\varepsilon _{{\rm p}}\right) ^{2}},  \label{FI}
\end{equation}%
represents the contribution of the pseudoscalar interaction. The
''triangle'' condition $p_{{\rm p}}+p_{l}>p_{{\rm n}}$, required by the
step-function, is necessary for conservation of the total momentum in the
reaction and exhibits the threshold dependence on the proton concentration.

\section{Non-relativistic limit}

Under beta-equilibrium, the superthreshold proton fraction in the core of
neutron stars appears at large densities, when Fermi momenta of nucleons are
of the order of their effective mass. Therefore the ''triangle'' condition
is inconsistent with the non-relativistic limit. Nevertheless, it is
necessary to show how the expression given by Lattimer et al. \cite{Lat91}
can be formally obtained from Eq. (\ref{QMFA}). For this purpose we consider
the non-relativistic limit of Eq. (\ref{QMFA}) by neglecting the
''triangle'' condition.

The non-relativistic approximation is valid when $p_{{\rm n}},p_{{\rm p}%
},p_{l}\ll M^{\ast }$. However, the smallness of the particle momenta is not
enough to point out unambiguously the leading terms in Eq. (\ref{QMFA}). The
relative contributions of various terms depend also on electron abundance in
the medium. If the electron fraction is as large that $M^{\ast }\mu _{l}\gg
p_{{\rm n}}^{2}$ then we obtain the result of Lattimer et al. \cite{Lat91}: 
\begin{equation}
Q_{L}=\frac{457\pi }{10080}G_{F}^{2}C^{2}\left( C_{V}^{2}+3C_{A}^{2}\right)
T^{6}M^{\ast 2}\mu _{l}\Theta \left( p_{l}+p_{{\rm p}}-p_{{\rm n}}\right) ,
\label{QL}
\end{equation}%
widely used in simulations of neutron stars. However, when $p_{{\rm n}%
}^{2}\sim M^{\ast }\mu _{l}$, what actually takes place in the
non-relativistic nucleon matter under beta-equilibrium, we obtain%
\begin{equation}
Q_{nr}=\frac{457\pi }{10080}G_{F}^{2}C^{2}T^{6}\left[ \left(
C_{V}^{2}+3C_{A}^{2}\right) M^{\ast 2}\mu _{l}-\left(
C_{V}^{2}+C_{A}^{2}\right) M^{\ast }p_{{\rm n}}^{2}\right] \Theta \left(
p_{l}+p_{{\rm p}}-p_{{\rm n}}\right) .  \label{Qnr}
\end{equation}%
Here the additional term is due to the nucleon recoil. When the electron
(proton) fraction is small, this contribution is comparable to the terms of
Eq. (\ref{QL}).

\section{Efficiency of the relativistic approach}

Neutrino energy losses caused by the direct Urca on nucleons depend
essentially on the composition of beta-stable nuclear matter. Therefore, in
order to estimate the relativistic effects, we consider the model of nuclear
matter, which includes nucleon and hyperon degrees of freedom (See
Appendix). The parameters of the model are chosen as suggested in Ref. \cite%
{GM} to reproduce the nuclear matter equilibrium density, the binding energy
per nucleon, the symmetry energy, the compression modulus, and the nucleon
effective mass at saturation density $n_{0}=0.16$ $fm^{-3}$. The composition
of neutrino-free matter in beta equilibrium among nucleons, hyperons,
electrons and muons is shown in Fig. 1 versus the baryon number density $%
n_{b}$, in units of $n_{0}$ (left panel). The right panel represents the
results of self-consistent calculation of the effective baryon masses and
individual Fermi momenta of the baryons\footnote{%
In order to apply the non-relativistic Eq. (\ref{QL}) to relativistic
nucleons the authors \cite{PBPELK97} suggest replacing the effective Dirac
mass with the relativistic Fermi energy, $M_{{\rm n,p}}^{\ast }\rightarrow 
\sqrt{p_{{\rm n,p}}^{2}+M^{\ast 2}}$. The latter is termed as ''the
effective Landau mass''. This trick can not be justified from the
theoretical point of view. Firstly because the mass term and the energy term
in the Dirac equation are of different matrix structure. Secondly, by the
correct definition, the effective Landau mass has no relation to relativism
but is intended to take into account interactions between particles near the
Fermi surface. In this meaning the effective mass (\ref{Mef}) is actually
the effective Landau mass.}.

At number densities smaller than the saturation density the matter consists
mostly of neutrons with a small admixture of equal amounts of protons and
electrons, however, the proton fraction grows along with the matter density.
Near saturation density the electron chemical potential reaches the muon
mass, and the muon fraction appears also growing along with the matter
density together with the electron fraction up to $n_{b}\simeq 0.3$ $fm^{-3}$%
. Above this density of nuclear matter, constituents other than neutrons,
protons, electrons, and muons appear.\ The $\Sigma ^{-}$ hyperons appear
when their lowest energy state first lies below $\mu _{{\rm n}}+\mu _{l}$.
Due to the charge neutrality, increasing of the $\Sigma ^{-}$ fraction
suppresses the lepton abundance.

\vskip0.3cm \psfig{file=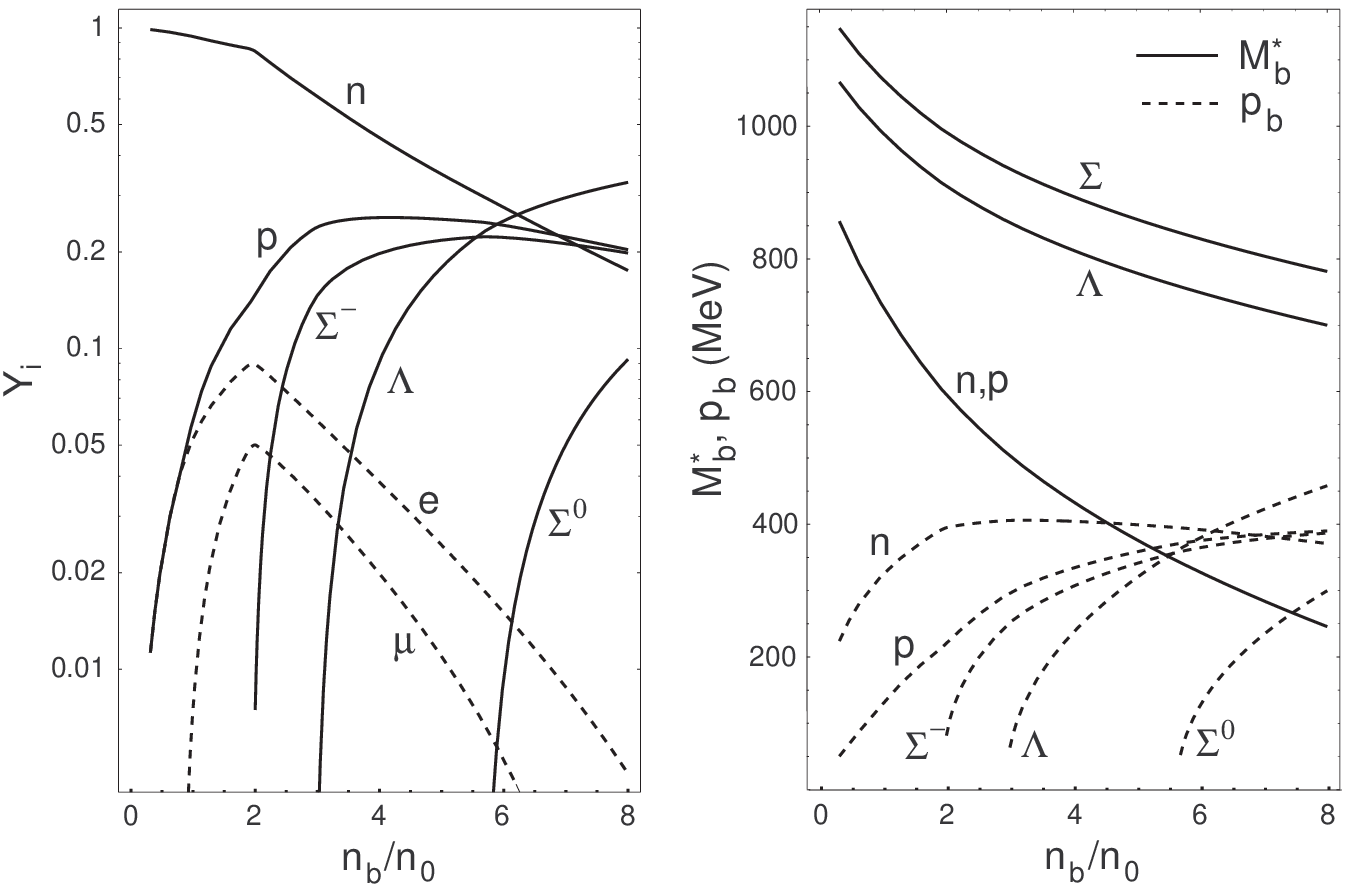}Fig. 1. The left panel shows individual
concentrations for matter in beta equilibrium among nucleons, hyperons,
electrons and muons as a function of the density ratio $n_{b}/n_{0}$. The
right panel represents the self-consistent effective baryon masses (solid
lines) and individual Fermi momenta of the baryons (dashed lines).

\vskip0.3cm

Above the density $n_{b}\simeq 0.48$ $fm^{-3}$, beta equilibrium requires
appearance of $\Lambda $ hyperons because the lowest energy state for a $%
\Lambda $ lies lower than $\mu _{{\rm n}}$. At densities larger than $%
n_{b}\simeq 0.91$ $fm^{-3}$ fraction of $\Sigma ^{0}$ hyperons also exists.

In Fig. 2 we compare the relativistic neutrino emissivity (\ref{QMFA}) with
the non-relativistic energy losses, as given by formula (\ref{QL}). Due to a
steady decrease of the effective mass of the nucleon the non-relativistic
formula (short-dashed curve) predicts a decreasing of the emissivity along
with the density increase. Appearance of hyperons in the system suppresses
the nucleon fractions and lepton abundance. Therefore at densities, where
the number of hyperons is comparable with the number of protons, the
relativistic emissivity (solid curve) reaches the maximum and then also has
a tendency to decrease. The relativistic emissivity, however, is found to be
substantially larger than that predicted in the non-relativistic approach.

To inspect the contribution of weak magnetism and the pseudoscalar
interaction we demonstrate two additional graphs. The long-dashed curve
demonstrates the energy losses obtained from Eq. (\ref{QMFA}) by formal
setting $\Phi =0$. This eliminates the pseudoscalar contribution. The
dot-dashed curve is obtained by formal replacing $\Phi =0$ and $C_{M}=0$,
which eliminates both the weak magnetism and pseudoscalar contributions. A
comparison of these curves demonstrates the weak magnetism effects. The
contribution of the pseudoscalar interaction can be observed by comparing
the total neutrino energy losses (solid curve) with the long-dashed curve,
which is calculated without this contribution. We see that, due to weak
magnetism effects, relativistic emissivities increase by approximately
40-50\%, while the pseudoscalar interaction only slightly suppresses the
energy losses, approximately by 5\%.

\vskip0.3cm \psfig{file=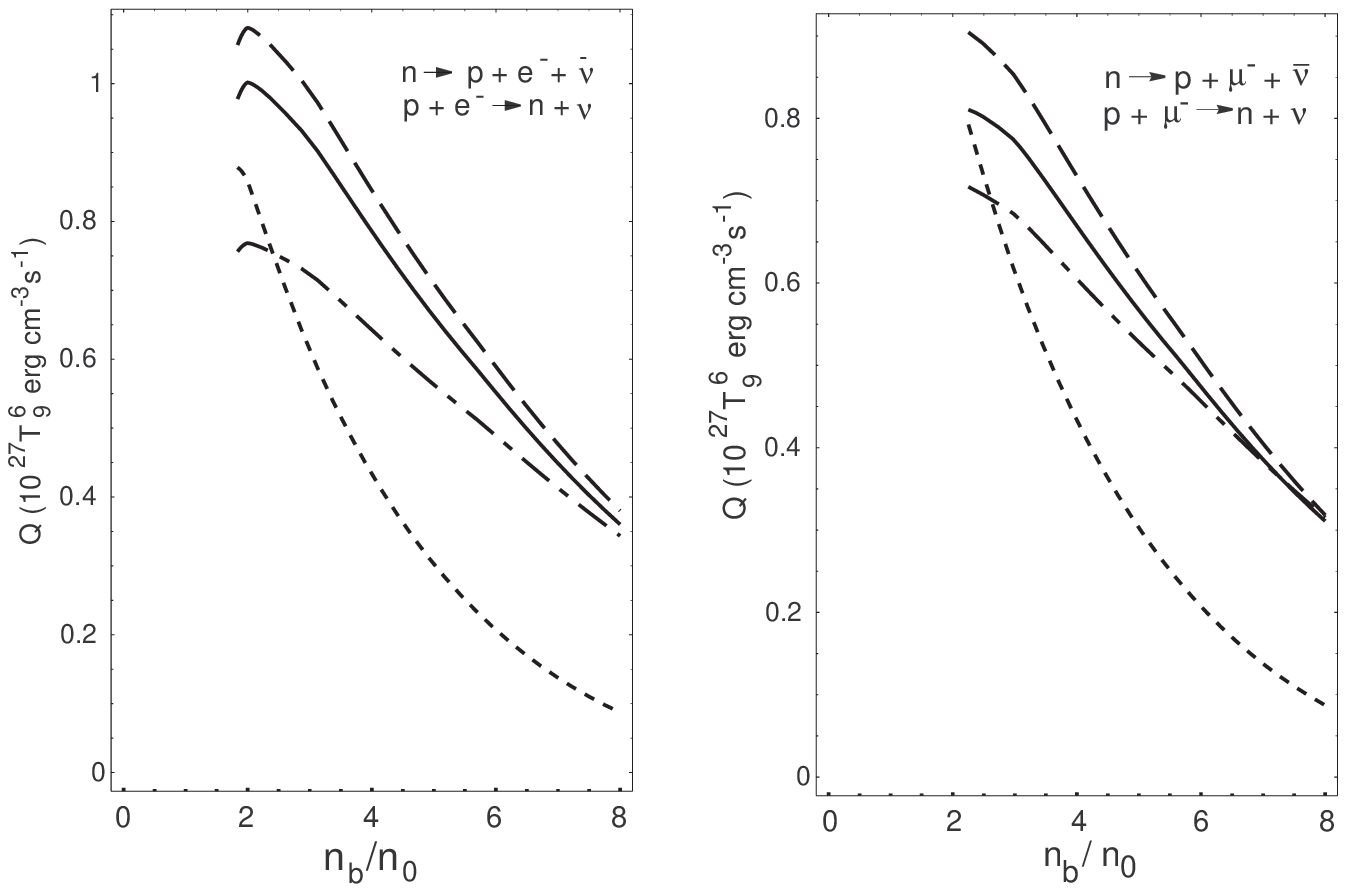}Fig. 2. The left panel shows the neutrino
emissivity of the direct Urca processes among nucleons and electrons. The
right panel is the same, but for the muon reactions. The emissivity is shown
versus the density ratio $n_{b}/n_{0}$ for the matter composition
represented in Fig. 1. The curves begin at the threshold density. Solid
curves represent the total relativistic emissivity, as given by Eq. (\ref%
{QMFA}). The short-dashed curves are the non-relativistic emissivity given
by Eq. (\ref{QL}). The dot-dashed curves show the relativistic emissivity
without the contributions of weak magnetism and pseudoscalar interaction,
and the long-dashed curves are the emissivity without the pseudoscalar
contribution. All the emissivities are given in units $10^{27}T_{9}^{6}$ $%
erg\,cm^{-3}s^{-1}$, where the temperature $T_{9}=T/10^{9}\,K$.
\vskip0.3cm

\section{Summary and Conclusion}

In the mean field approximation, we have studied the neutrino energy losses
caused by the direct Urca processes on nucleons in the degenerate baryon
matter under beta equilibrium. We have shown that the direct Urca processes
in a superdense matter of neutron star cores are kinematically allowed only
due to isovector mesons, which differently interact with protons and
neutrons. By creating the energy gap between proton and neutron spectrums
the isovector mesons support a time-like total momentum transfer from the
nucleon, as required by kinematics of the reaction. In the mean field
approximation, we derived the matrix element of the nucleon transition
current, which is found to be a function of the space-like kinetic momentum
transfer. We have calculated the neutrino energy losses caused by the direct
Urca processes on nucleons. Our Eq. (\ref{QMFA}) for neutrino energy losses
exactly incorporates the effects of nucleon recoil, parity violation, weak
magnetism, and pseudoscalar interaction. To quantify the relativistic
effects we consider a self-consistent relativistic model, widely used in the
theory of relativistic nuclear matter. The relativistic energy losses are up
to four times larger than those given by the non-relativistic approach. In
our analysis, we pay special attention to the effects of weak magnetism and
pseudoscalar interaction in the neutrino energy losses. We found that, due
to weak magnetism effects, relativistic emissivities increase by
approximately 40-50\%, while the pseudoscalar interaction only slightly
suppresses the energy losses, approximately by 5\%. The mean field Eq. (\ref%
{QMFA}) may be considered as a starting point for studying of the
correlation effects.

\section*{Acknowledgements}

This work was partially supported by the Russian Foundation for Fundamental
Research Grant 00-02-16271

\appendix

\section{Field theoretical models of nuclear matter}

We employ a self-consistent relativistic model of nuclear matter in which
baryons, $B=n,p,\Sigma ^{-},\Sigma ^{0},\Sigma ^{+},\Lambda $, interact via
exchange of $\sigma $, $\omega $, and $\rho $ mesons \cite{Serot}. In
principle, the pion fields should be also included in the model. However,
the expectation value of the pion field equals zero, giving no contribution
to the mean fields. Therefore, only non-redundant terms are exhibited in the
Lagrangian density%
\begin{eqnarray}
{\cal L} &=&\sum_{B}\bar{B}\left[ \gamma _{\mu }\left( i\partial ^{\mu
}-g_{\omega B}\omega ^{\mu }-\frac{1}{2}g_{\rho B}{\vec b}^{\mu }\cdot {\vec %
\tau }\right) -\left( M_{B}-g_{\sigma B}\sigma \right) \right] B  \nonumber
\\
&&-\frac{1}{4}F_{\mu \nu }F^{\mu \nu }+\frac{1}{2}m_{\omega }^{2}\omega
_{\mu }\omega ^{\mu }-\frac{1}{4}{\vec B}_{\mu \nu }{\vec B}^{\mu \nu }+\frac{1%
}{2}m_{\rho }^{2}{\vec b}_{\mu }{\vec b}^{\mu }  \nonumber \\
&&+\frac{1}{2}\left( \partial _{\mu }\sigma \partial ^{\mu }\sigma
-m_{\sigma }^{2}\sigma ^{2}\right) -U\left( \sigma \right) +\bar{l}\left(
i\gamma _{\mu }\partial ^{\mu }-m_{l}\right) l,  \label{Lagr}
\end{eqnarray}%
which includes the interaction of baryon fields $B$ with a scalar field $%
\sigma $, a vector field $\omega _{\mu }$ and an isovector field ${\vec b}%
_{\mu }$ of $\rho $-meson. In the above, $B$ are the Dirac spinor fields for
baryons, ${\vec b}_{\mu }$ is the isovector field of $\rho $-meson. We denote
as ${\vec \tau }$ the isospin operator, which acts on the baryons of the bare
mass $M_{B}$. The leptons are represented only by electrons and muons, $%
l=e^{-},\mu ^{-}$, which are included in the model as noninteracting
particles. The field strength tensors for the $\omega $ and $\rho $ mesons
are $F_{\mu \nu }=\partial _{\mu }\omega _{\nu }-\partial _{\nu }\omega
_{\mu }$ and ${\vec B}_{\mu \nu }=\partial _{\mu }{\vec b}_{\nu }-\partial
_{\nu }{\vec b}_{\mu }$, respectively. The potential $U\left( \sigma \right) $
represents the self-interactions of the scalar field and is taken to be of
the form 
\begin{equation}
U\left( \sigma \right) =\frac{1}{3}bM\left( g_{\sigma N}\sigma \right) ^{3}+%
\frac{1}{4}c\left( g_{\sigma N}\sigma \right) ^{4}.  \label{selfint}
\end{equation}

In what follows we consider the mean field approximation widely used in the
theory of relativistic nuclear matter. In this approximation, the meson
fields are replaced with their expectation values%
\begin{equation}
\sigma \rightarrow \left\langle \sigma \right\rangle \equiv \sigma _{0},\,\
\ \omega ^{\mu }\rightarrow \left\langle \omega ^{\mu }\right\rangle \equiv
\omega _{0}\delta _{\mu 0},\,\ \ \ {\vec b}^{\mu }\rightarrow \left\langle 
{\vec b}^{\mu }\right\rangle \equiv \left( 0,0,\rho _{0}\right) \delta _{\mu
0}.  \label{avfield}
\end{equation}%
In this case only the baryon fields must be quantized. This procedure yields
the following linear Dirac equation for the baryon field 
\begin{equation}
\left( i\partial _{\mu }\gamma ^{\mu }-g_{\omega B}\gamma ^{0}\omega _{0}-%
\frac{1}{2}g_{\rho B}\gamma ^{0}\rho _{0}\tau _{3}-\left( M-g_{\sigma
B}\sigma _{0}\right) \right) B\left( x\right) =0,
\end{equation}%
The effective baryon mass is $M_{B}^{\ast }=M_{B}-g_{\sigma B}\sigma _{0}$,
and the single-particle energies and self-consistent potentials are given by 
\begin{equation}
U_{B}=g_{\omega B}\omega _{0}+g_{\rho B}t_{3B}\rho _{0}.  \label{U}
\end{equation}%
Here $\omega _{0}$, $b_{0}$, and $\sigma _{0}$ are, respestively, nonzero
average values of the $\omega $-, $\rho $-, and $\sigma $-meson fields, and $%
g_{\omega B}$, $g_{\rho B}$, and $g_{\sigma B}$ are the strong interaction
couplings of the baryons to different meson fields, and $t_{3B}$ is the
third component of isospin for the baryons. The number density for baryon
species $B$ is related to the corresponding Fermi momentum as 
\begin{equation}
n_{B}=\frac{p_{B}^{3}}{3\pi ^{2}}.  \label{numden}
\end{equation}%
Thus, the average meson fields are solutions of the following equations 
\begin{eqnarray}
m_{\omega }^{2}\omega _{0} &=&\sum_{B}g_{\omega B}n_{B},  \nonumber \\
m_{\rho }^{2}\rho _{0} &=&\sum_{B}g_{\rho B}t_{3B}n_{B},  \label{meson} \\
m_{\sigma }^{2}\sigma _{0} &=&-\frac{dU\left( \sigma \right) }{d\sigma }%
+\sum_{B}\frac{g_{\sigma B}}{2\pi ^{2}}\left( M_{B}^{\ast }p_{B}\sqrt{%
p_{B}^{2}+M_{B}^{\ast 2}}-M_{B}^{\ast 3}\arcsin \frac{p_{B}}{M_{B}^{\ast }}%
\right) .  \nonumber
\end{eqnarray}%
For a fixed total baryon number density 
\begin{equation}
n_{b}=\sum_{B}n_{B},  \label{ntot}
\end{equation}%
the additional conditions needed to obtain a solution are given by the
charge neutrality requirement%
\begin{equation}
n_{e}+n_{\mu }+n_{\Sigma ^{-}}=n_{p}+n_{\Sigma ^{+}}  \label{neutr}
\end{equation}%
and the beta equilibrium relations 
\begin{eqnarray}
\mu _{n} &=&\mu _{p}+\mu _{e},\,\ \ \ \mu _{\mu }=\mu _{e},\ \ \ \ \mu
_{\Sigma ^{-}}=\mu _{n}+\mu _{e},  \nonumber \\
\mu _{\Sigma ^{0}} &=&\mu _{\Lambda }=\mu _{n},\ \ \ \ \mu _{\Sigma
^{+}}=\mu _{p}  \label{beta}
\end{eqnarray}%
For nucleons, the strong interaction constants \cite{GM} 
\begin{eqnarray}
\frac{g_{\sigma N}}{m_{\sigma }} &=&3.15\ fm,\ \ \ \frac{g_{\omega N}}{%
m_{\omega }}=2.20\ fm,\ \ \ \frac{g_{\rho N}}{m_{\rho }}=2.19\ fm,  \nonumber
\\
b &=&0.008659,\ \ \ c=-0.002421,  \label{strong}
\end{eqnarray}%
with $m_{\omega }=783$ MeV,$\ \ \ \ \ m_{\rho }=770$ MeV,$\ \ \ \ m_{\sigma
}=520$ MeV, are determined by reproducing the nuclear matter equilibrium
density $n_{0}=0.16$ $fm^{-3}$, and the binding energy per nucleon ($\sim
16\ $MeV), the symmetry energy ($\sim 30-35\ $MeV), the compression modulus
( $200\ $MeV$\leq K_{0}\leq 300\ $MeV), and the nucleon effective mass $%
M^{\ast }=\left( 0.6-0.7\right) \times 939\ $MeV at $n_{0}$. For hyperons we
take $g_{\sigma H}=0.6g_{\sigma N},\,g_{\rho H}=0.6g_{\rho N},\,g_{\omega
H}=0.658g_{\omega N}$.

\end{document}